\documentclass[sigconf]{acmart}
\usepackage[utf8]{inputenc}
\usepackage{multirow}
\usepackage{makecell}
\usepackage{xcolor}
\usepackage{xspace}
\usepackage{hyperref}
\AtBeginDocument{%
  \providecommand\BibTeX{{%
    \normalfont B\kern-0.5em{\scshape i\kern-0.25em b}\kern-0.8em\TeX}}}
    
\acmJournal{TOG}

\copyrightyear{2023} 
\acmYear{2023} 
\setcopyright{acmlicensed}\acmConference[SIGGRAPH '23 Conference Proceedings]{Special Interest Group on Computer Graphics and Interactive Techniques Conference Conference Proceedings}{August 6--10, 2023}{Los Angeles, CA, USA}
\acmBooktitle{Special Interest Group on Computer Graphics and Interactive Techniques Conference Conference Proceedings (SIGGRAPH '23 Conference Proceedings), August 6--10, 2023, Los Angeles, CA, USA}
\acmPrice{15.00}
\acmDOI{10.1145/3588432.3591511}
\acmISBN{979-8-4007-0159-7/23/08}

\UseRawInputEncoding
\newcommand\ie{\textit{i.e.,} }
\newcommand\etal{\textit{et al.}}

\newcommand\tb[1]{\textbf{#1}}

\newcommand\rb[1]{#1}
\newcommand{\ignorethis}[1]{}

\begin{document}

\title{In the Blink of an Eye: Event-based Emotion Recognition}

\author{Haiwei Zhang}\authornotemark[1]
\affiliation{%
  \authornote{Equal contribution.}
  \institution{Dalian University of Technology}
  \city{Dalian}
  \country{China}
}
\orcid{0009-0007-8823-0554}
\email{haiweizhang32009182@mail.dlut.edu.cn}
  
\author{Jiqing Zhang}
\authornotemark[1]
\affiliation{%
  \institution{Dalian University of Technology}
  \city{Dalian}
  \country{China}
}
\orcid{0000-0002-0061-5465}
\email{jqz@mail.dlut.edu.cn}

\author{Bo Dong}
\authornotemark[2]
\affiliation{%
\authornote{Corresponding authors.}
  \institution{Princeton University}
  \city{Princeton}
  \country{USA}}
\orcid{0000-0001-9189-9506}
\email{bo.dong@princeton.edu}

\author{Pieter Peers}
\affiliation{%
  \institution{College of William \& Mary}
  \city{Williamsburg}
  \country{USA}}
  \orcid{0000-0001-7621-9808}
  \email{ppeers@siggraph.org}

\author{Wenwei Wu}
\affiliation{%
 \institution{Dalian University of Technology}
 \city{Dalian}
 \country{China}}
 \orcid{0009-0000-4241-0049}
 \email{wuwenwei0206@mail.dlut.edu.cn}

\author{Xiaopeng Wei}
\authornotemark[2]
\affiliation{%
  \institution{Dalian University of Technology}
  \city{Dalian}
  \country{China}}
  \orcid{0000-0002-8497-611X}
 \email{xpwei@dlut.edu.cn}

\author{Felix Heide}
\affiliation{%
  \institution{Princeton University}
  \city{Princeton}
  \country{USA}}
  \orcid{0000-0002-8054-9823}
  \email{fheide@cs.princeton.edu}

\author{Xin Yang}
\authornotemark[2]
\affiliation{
  \institution{Key Laboratory of Social Computing and Cognitive Intelligence of Ministry of Education, Dalian University of Technology}
  \city{Dalian}
  \country{China}}
  \orcid{0000-0002-8046-722X}
\email{xinyang@dlut.edu.cn}

\renewcommand{\shortauthors}{Zhang, et al.}

\begin{abstract}
We introduce a wearable single-eye emotion recognition device and a real-time approach to recognizing emotions from partial observations of an emotion that is robust to changes in lighting conditions. At the heart of our method is a bio-inspired event-based camera setup and a newly designed lightweight Spiking Eye Emotion Network (SEEN). Compared to conventional cameras, event-based cameras offer a higher dynamic range (up to 140 dB vs. 80 dB) and a higher temporal resolution (in the order of $\mu$s vs. 10s of $m$s). Thus, the captured events can encode rich temporal cues under challenging lighting conditions. However, these events lack texture information, posing problems in decoding temporal information effectively. SEEN tackles this issue from two different perspectives. First, we adopt convolutional spiking layers to take advantage of the spiking neural network's ability to decode pertinent temporal information. Second, SEEN learns to extract essential spatial cues from corresponding intensity frames and leverages a novel  weight-copy scheme to  convey spatial attention to the convolutional spiking layers during training and inference. We extensively validate and demonstrate the effectiveness of our approach on a specially collected Single-eye Event-based Emotion (SEE) dataset. To the best of our knowledge, our method is the first eye-based emotion recognition method that leverages event-based cameras and spiking neural networks.
\end{abstract} 


\begin{teaserfigure}
  \includegraphics[width=\textwidth]{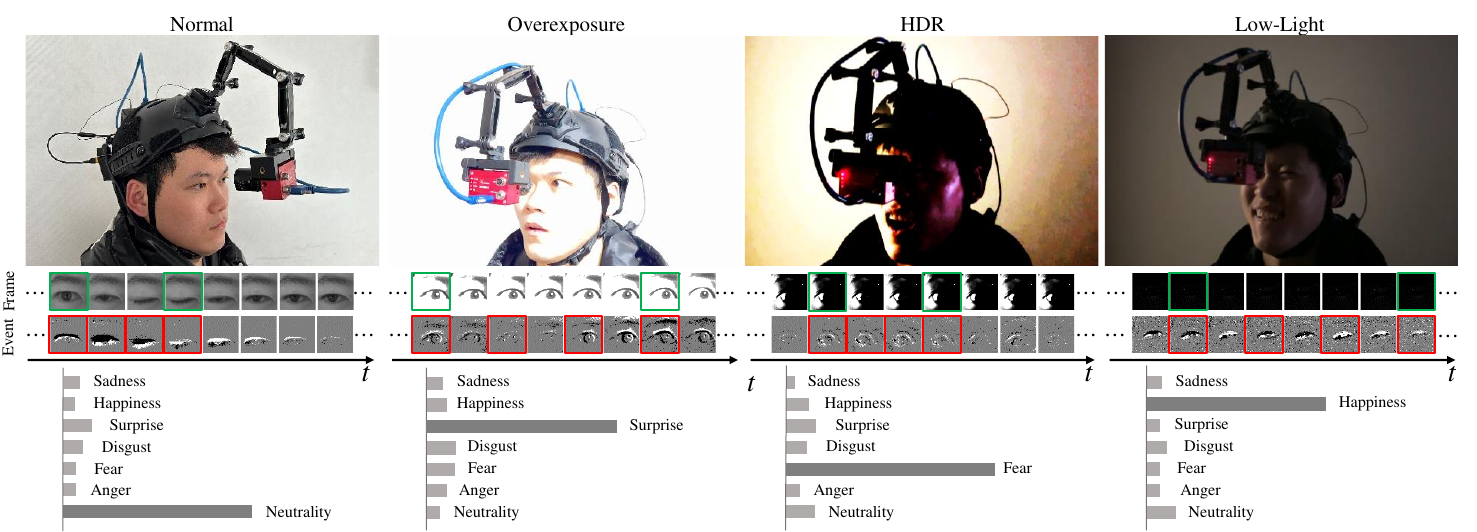}
  \caption{
  Demonstration of a wearable single-eye emotion recognition prototype system consisting with a bio-inspired event-based camera (DAVIS346) and a low-power NVIDIA Jetson TX2 computing device. Event-based cameras simultaneously provide intensity and corresponding events, which we input to a newly designed lightweight Spiking Eye Emotion Network (SEEN) to effectively extract and combine spatial and temporal cues for emotion recognition. Given a sequence, SEEN takes the start and end intensity frames (green boxes) along with $n$ intermediate event frames (red boxes) as input. Our prototype system consistently recognizes emotions based on single-eye areas under different lighting conditions at $30$ FPS.
  } 
  \Description{teaser figure}
  \label{fig:teaser}
\end{teaserfigure}


\maketitle
\UseRawInputEncoding
\section{introduction}

Real-time emotion recognition in uncontrolled environments is a challenging problem that forms the cornerstone of many \textit{in-the-wild} human-centered interactive computer graphics experiences such as interactive storytelling that adapts to the users emotions, and emotion-aware virtual avatars. Predicting emotions from regular RGB video streams is a challenging and ambiguous endeavor; informative spatial and temporal emotive cues can be adversely affected by head pose and partial occlusions. To help classify emotions in RGB video frames, existing facial emotion recognition models build on complex CNN-based models such as ResNet 50~\cite{deng2020mimamo}, Transformer~\cite{zhao2021former},  and Incepti\-on-based methods~\cite{hickson2019eyemotion}. Robustly handling varying lighting conditions and rapid user movements further complicates emotion recognition, and existing methods rely on cumbersome large network enhancement modules~\cite{zhao2021former} or impose active IR lighting~\cite{wu2020emo}.  Despite all these innovations, emotion recognition from RGB video streams remains difficult and fragile.

In this paper, we introduce a novel wearable emotion recognition prototype in which a bio-inspired event-based camera (DAVIS\-346) is affixed in front of a user's right eye. An event-based camera can provide more robust temporal cues for emotion recognition under adverse lighting conditions as it offers a higher dynamic range (up to $140$ dB vs. $80$ dB) and a higher temporal resolution (in the order of $\mu$s vs. $10$s of $ms$) than a conventional camera. Even though this setup provides a stable fixed perspective of a right eye and it can robustly handle various lighting conditions, estimating emotion from a single eye still poses unique challenges. 

A key issue is that event-based cameras do not capture texture information effectively (see \autoref{fig:teaser}). These spatial features are not only essential for emotion recognition but also important for inferring more informative temporal features. For example, while pupil motion and blinking are dominant temporal cues, they are less informative for emotion classification. In contrast, the subtle movements related to the facial units, such as raising the outer brow and squinting, are stronger cues for eye-based emotion recognition.

To address these challenges, we devise a lightweight SEEN, which combines the best from both events and intensity frames to \textit{guide} emotion recognition from asynchronous events with spatial texture cues from corresponding intensity frames. In particular, SEEN consists of a spatial feature extractor and a temporal feature extractor that partially share the same convolutional architecture. During training, the shared convolutional parts are only learned in the spatial feature extractor, and the updated weights are copied to the temporal feature extractor. Consequently, spatial attention can be effectively conveyed to the temporal decoding process. As such, the temporal feature extractor learns to associate spatial and temporal features, resulting in a consistent emotion classification. 

To train our lightweight Spiking Eye Emotion Network (SEEN) and to stimulate research in event-based single-eye emotion recognition, we introduce a new Single-eye Event-based Emotion (SEE) dataset. We validate our approach on the SEE dataset and demonstrate state-of-the-art emotion recognition under different challenging lighting conditions, outperforming the runner-up method by a significant margin, $4.8\%$ and $4.6\%$ in WAR and UAR, respectively. The prototype system with an NVIDIA Jetson TX2 operates at $30$ FPS in real-world testing scenarios. 

Specifically, our work makes the following contributions: 
\begin{itemize}
	\item a novel real-time emotion recognition method based on event camera measurements and a spiking neural network suited for in-the-wild deployment;
	\item a weight-copy training scheme to enforce learned weights awareness of both spatial and temporal cues; and
	\item the first publicly available single-eye emotion dataset containing both intensity frames and corresponding raw events, captured under four different lighting conditions.
\end{itemize}

\textit{Limitations.}
SEEN partially relies on spatial features extracted from intensity frames, which can be adversely affected by extremely degraded lighting conditions, resulting in a significant performance drop. While our method robustly handles most lighting conditions effectively, as evidenced by our experimental results, further improving robustness by solely leveraging events forms an exciting avenue for future research in eye-based emotion recognition. 

The code and dataset are available on~\href{https://github.com/zhanghaiwei1234/Single-eye-Emotion-Recognition}{\underline{github}}.

\UseRawInputEncoding
\section{Related Work}

We focus our discussion on related work in emotion recognition on measuring emotions (wearable emotion sensing systems) and recognition (facial emotion recognition).

\textit{Wearable Emotion Sensing Systems.}
Emotions impact the human body in subtle ways. However, not all of these signals are equally robust indicators of emotional state, and not all are easily measured.  Various bio-signals have been investigated for convenient measurement of indicators of emotional state.  Long-term heart rate variability (HRV) has been shown to strongly correlate with emotional patterns~\cite{10.1145/3328911, doi:10.1037/1089-2680.10.3.229}.  Similarly, brain activity recorded by electroencephalogram (EEG) sensors also correlates to different emotions~\cite{Li2018EmotionRF, 10.3389/fnsys.2020.00043}.  Inspired by human perception of emotions, Electromyogram (EMG) measurements of facial muscle contractions~\cite{Lucero1999AMO} map to emotions, making wearable emotional detection devices possible~\cite{6778017}.   A disadvantage of these methods is that they require the sensors to make \textit{direct skin contact}, dramatically restricting freedom of activity.  Furthermore, due to the displacement of sensors and muscular cross-talk during movement, the results can be of low reliability. An alternative to contact-based measurement is pupillometry, \ie the measurement of pupil size and reactivity, as a potential indicator of emotion~\cite{Pupillometry:2018:02, 9097620}.  However, pupilometry requires expensive equipment, and the reliability is significantly impacted by ambient lighting~\cite{Couret2019TheEO}.  Similar to pupilometry, our method also focuses on the eye as an indicator of emotional state.  However, in contrast to prior work, we employ an event-based camera that does not require direct skin contact and which can operate in challenging lighting conditions.

\textit{Facial Emotion Recognition.}
Facial emotion recognition has received significant attention in computer graphics and computer vision, with applications ranging from driving facial expressions~\cite{hickson2019eyemotion, 10.1145/3528233.3530745} to facial reenactment for efficient social interactions~\cite{10.1145/2820903.2820910, 10.1145/2766939}. A significant portion of prior work in facial emotion recognition requires observations of the entire face, and several methods have been introduced for effective facial feature learning~\cite{xue2021transfer, ruan2021feature}, dealing with uncertainties in facial expression data~\cite{zhang2021relative}, handing partial occlusions~\cite{georgescu2019recognizing, houshmand2020facial}, and exploiting temporal cues~\cite{sanchez2021affective, deng2020mimamo}.  Combinations with other modalities such as contextual information~\cite{lee2019context} and depth~\cite{lee2020multi} have also been explored to further improve facial recognition accuracy. 

However, observing the entire face is not feasible in many practical situations.  Alternatively, several methods focus on the eye area only for emotion recognition.  Hickson~\etal~\shortcite{hickson2019eyemotion} infer emotional expressions based on images of both eyes captured with an infrared gaze-tracking camera inside a virtual reality headset.
Wu~\etal~\shortcite{wu2020emo} rely on infrared single-eye observations to reduce camera synchronization and data bandwidth issues when monitoring both eyes. Both systems require a personalized initialization procedure; Hickson~\etal~ require a personalized neutral image, and Wu~\etal~ require a reference feature vector of each emotion. The need for a personalized setup  makes these systems intrusive and non-transparent to the user and could raise privacy concerns. Furthermore, neither system leverages temporal cues, which are essential for robust emotion recognition~\cite{sanchez2021affective}. Our approach does not require personalization, and it leverages temporal and spatial cues to improve emotion recognition accuracy.

\UseRawInputEncoding
\section{background}
\label{sec:background}

Before detailing our method, we first review work related to the two key components of our event-based emotion recognition method: event-based cameras and spiking neural networks.  

\textit{Event-based Cameras.}
An event-based camera differs from a conventional camera in that it does not measure pixel intensities, but instead, an event-based camera records asynchronous (log-encoded) per-pixel brightness changes~\cite{gehrig2021combining, 9138762}. Event-based cameras offer a significantly higher dynamic range (up to 140 dB) and a higher temporal resolution (in the order of $\mu$s) than conventional cameras.   Each event $e$ is characterized by three pieces of information: the pixel location, $(x, y)$; the event triggering time, $t$; and a polarity, $p \in \{-1, 1\}$ which reflects the direction of the brightness change. Formally, a set of $N$ events can be defined as: 

\begin{equation} 
\label{eq:event_set}
\mathcal{E} = \{e_k\}_{k=1}^N = \{[x_k, y_k, t_k, p_k]\}_{k=1}^N.
\end{equation}

Under static lighting, a stationary event-based camera only records scene motion, and events are typically triggered by moving edges (\textit{e.g.}, object contours, and texture boundaries). Since the events predominately stem from the motion of edges, the measured events are inherently sparse and devoid of texture information.  Furthermore, since the captured events are triggered asynchronously, events are incompatible with CNN-based architectures.  Instead, events are aggregated into a frame or grid-based representation \cite{ gehrig2019end,lagorce2016hots, maqueda2018event,wang2021event} before neural processing. In our implementation, we adopt the aggregation algorithm of Zhang~\etal~\shortcite{zhang2021object}, which currently offers the highest performance for single object tracking  under normal and degraded conditions. We refer to the Supplementary Material for additional details.

\textit{Spiking Neural Network (SNN).}
Spiking neural networks (SNNs) closely mimic biological information processes.  An SNN incorporates the concept of time and only exchanges information (\ie spike) when a  \textit{membrane potential}  exceeds some potential threshold.  Mathematically an SNN neuron simulates the properties of a cell in a nervous system with varying degrees of detail, which models three states of a biological neuron: rest, depolarization, and hyperpolarization~\cite{ding2022biologically}.  When a neuron is at rest, its membrane potential remains constant; typically set to $0$. When not at rest, the change in the membrane potential can either decrease or increase.  An increase in membrane potential is called depolarization. In contrast, hyperpolarization describes a reduction in membrane potential. When a membrane potential is higher than a potential threshold, an action potential, \ie spike, is triggered, which for an SNN is a binary value. We refer the interested reader to Ding \etal~\shortcite{ding2022biologically} for an in-depth discussion of these concepts. 

In this paper, we use the \textit{leaky integrate-and-fire (LIF)} spiking neuron model~\cite{gerstner2002spiking}, one of the most widely used spiking models. When a LIF neuron receives spikes from other neurons, the spikes are scaled accordingly based on learned synaptic weights. Depolarization is achieved by summing over all the scaled spikes. A decay function over time is used to drive the potential membrane to hyperpolarization. We refer to the Supplemental Material for a detailed formal definition of LIF.

\UseRawInputEncoding
\begin{figure*}[th!]
  \centering
  \includegraphics[width=\linewidth]{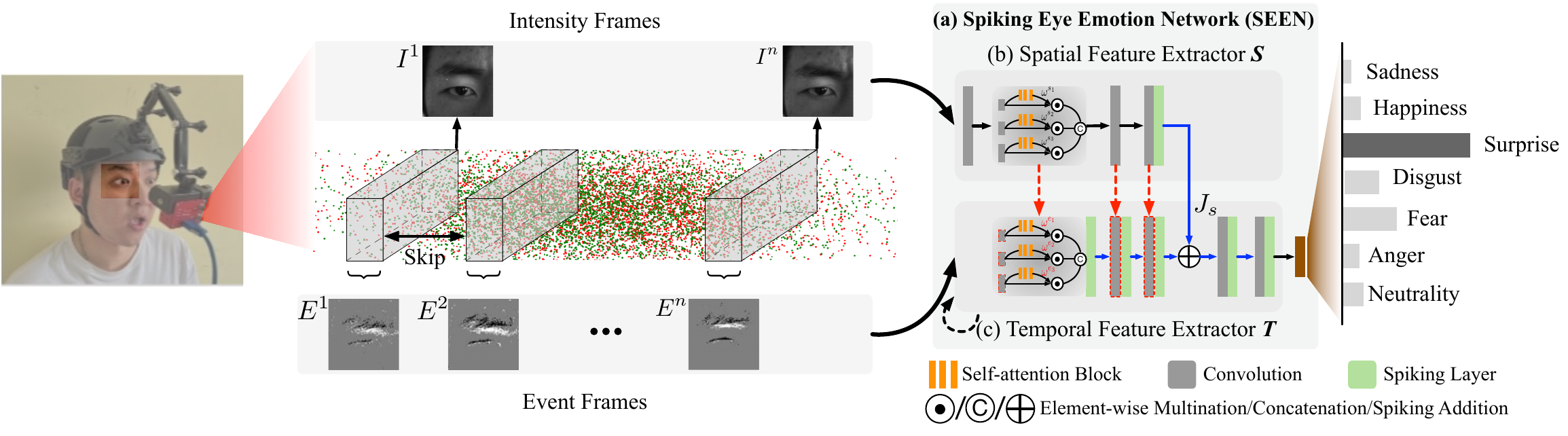}
  \caption{
  Our Spiking Eye Emotion Network (a) leverages a CNN-SNN-based temporal feature extractor, $T$, (c) to process $n$ accumulated event frames, \ie $E^i$, in time order sequentially. During the process, based on two intensity frames, $I^1$ and $I^n$, the spatial feature extractor, $S$, (b) relies on a multiscale self-attention module to extract spatial features, $J_s$, which are combined with temporal cues to estimate emotions. The convolutional blocks before spiking-addition operator in the temporal feature extractor $T$ fail to properly train due to lack of texture information in the event frames. Instead, we copy the updated weights from the corresponding blocks of the spatial feature extractor $S$. During inference, the attention weights are also copied directly from $S$ to $T$ to increase inference speed. The copying operations are marked by the red dashed arrows.}
  \label{fig:SEEN_pipeline}
\end{figure*}

\section{Spiking Eye Emotion Network (SEEN)}
\label{sec:SEEN}

Existing facial emotion recognition methods typically only identify the \textit{``peak''} states of emotions~\cite{hickson2019eyemotion} or a single emotion state over a whole sequence~\cite{zhao2021former}, making these methods unsuitable for applications that also require a robust estimate of the in-between states.  We introduce a lightweight Spiking Eye Emotion Network (SEEN) that is able to effectively recognize emotions from various states of emotions.   

Instead of only memorizing the peak phase of an individual's facial emotion, SEEN is designed to leverage temporal cues to distinguish different phases of emotions using sparse events input captured with an event-based camera (DAVIS346 camera). Compared to a conventional camera, an event-based camera has a number of advantages: it is more sensitive to motion, less sensitive to ambient lighting, and it offers a high dynamic range. Hence, an event-based camera is capable of providing stable temporal information under different lighting conditions. While this makes event-based cameras, in theory, an attractive input modality for motion-based measurements, in practice, a major drawback of existing event-based cameras is that the recorded events are noisy and lack texture information. We address this drawback with a hybrid system that leverages both spatial cues together with conventional intensity frames to guide temporal feature extraction during training and inference.  Most commercial event-based cameras are capable of simultaneously capturing both intensity frames and events through spatially-multiplexed sensing.

\subsection{SEEN Architecture}

As illustrated in \autoref{fig:SEEN_pipeline}(a), at its core, the architecture of SEEN consists of a spatial feature extractor, $S$ (described in detail in~\autoref{sec:S}), and a temporal feature extractor, $T$ (detailed in~\autoref{sec:T}). Given two intensity frames, $I^1$ and $I^n$, SEEN interpolates the asynchronous captured events between both intensity frames in $n$ synchronous event frames. Next, the spatial feature extractor $S$ distills spatial cues from the intensity frames $I^1$ and $I^n$, and the temporal feature extractor $T$ processes each of the $n$ event frames sequentially in time order. Finally, the temporal features and the spatial cues are then combined to predict $n$ emotion scores. The final predicted emotion is based on the average of the $n$ scores. The core component of the temporal feature extractor $T$ is the SNN layers that make decisions based on membrane potentials to remember temporal information from previous event frames. Unlike RNNs~\cite{kag2021time, nah2019recurrent}, SNNs can effectively learn temporal dependencies of arbitrary length without any special treatment.

\subsection{Spatial Feature Extractor $S$}
\label{sec:S}

To make spatial feature extraction independent from the intensity sequence length, we only use the first and last frames of a sequence as the input to the spatial feature extractor, thereby fixing the input size regardless of the sequence length, \ie two frames. The spatial feature extractor $S$ (\autoref{fig:SEEN_pipeline}(b)) leverages a multiscale self-attention perception module, $\Omega$, to obtain discriminative features from different-sized neighborhoods. The extracted spatial features are then transferred into the spiking format, $J_s$, via a spiking layer, which is subsequently combined with temporal features to enhance feature discrimination. Formally, the spatial feature extractor can be defined as:
\begin{align}
    J_s & = \Phi^1(F_s), \\
    F_s &= C_3(C_3(\Omega_{(3,5,7)}(l_s))), \label{eq:S_fs} \\
    \Omega_{(x_1,...,x_n)}(\cdot) & := C_1([\omega_{(x_1,...,x_n)}^{s_1} C_{x_1}(\cdot),...,\omega_{(x_1,...,x_n)}^{s_n} C_{x_n}(\cdot)]),\label{eq:S_Omega}\\
    \omega_{(x_1,..., x_n)}^{s_i} &= \sigma\left(\left<\Upsilon(C_{x_1}(l_s)),...\Upsilon(C_{x_n}(l_s))\right>\right)_i,\label{eq:omega}\\
    \Upsilon(\cdot) &:= C_1(\mathcal{BR}(C_1(\mathcal{A}(\cdot)))),\label{eq:Upsilon}\\
    l_s &= C_1([I^1, I^n]),\label{eq:l_s}
\end{align}
where $[\cdot]$ and $\left<\cdot\right>$ indicate channel-wise concatenation and a vector, respectively; $C_i$ and $\sigma$ denote an $i \times i$ convolution layer and a softmax function, respectively;  $\mathcal{A}$ denotes an adaptive pooling layer; $\mathcal{BR}$ is a fused batch normalization layer with a ReLU activation function; $\Phi^t$ is a spiking layer that keeps membrane potential from the previous time step, $t-1$. The initial membrane potential, \ie $t = 0$, is set to $0$ (see \autoref{eq:SNN_layer}).

\subsection{Temporal Feature Extractor $T$}
\label{sec:T}

The basis building blocks of the temporal feature extractor $T$ are SNN layers. An SNN neuron outputs signals based on a membrane potential accumulation, decay, and reset mechanisms to capture the temporal trends in an input sequence \cite{ding2022biologically}. 
When the membrane potential exceeds a threshold, an action potential (\ie spike) is triggered 
and the membrane potential is reset. The trigger process itself is non-dif\-fer\-en\-tia\-ble, prohibiting training via conventional stochastic gradient descent optimization methods. Instead, we adopt spatio-temporal backpropagation (STBP) along with a CNN-SNN layer \cite{wu2018spatio} to circumvent this issue. This CNN-SNN layer employs a CNN-based layer for the aggregation process and a LIF-based SNN neuron \cite{gerstner2002spiking} for managing the potential decay and reset processes.  This modification takes advantage of CNN-based layers that enable learning of diverse accumulation strategies, resulting in more effective SNN neurons in the temporal domain.

\textit{Intensity Attention-Guided Temporal Features.} 
Purely relying on events does not yield a robust solution due to the lack of reliable texture information in the event domain. We, therefore, leverage spatial features from $S$ to inject rich texture cues. 
\autoref{fig:SEEN_pipeline}(c) illustrates the architecture of the temporal extractor $T$.

The feature extractor $T$ takes $n$ event frames, $E^1$ to $E^n$, as input and processes each frame sequentially in time order. Formally, given the spatial feature $J_s$, the temporal feature extraction of $E^t$ is defined by:
\begin{align}
    O^t &= \mathcal{M}(\Gamma (\Gamma(J^t_c))), \label{eq:T_O}\\
    J^t_c &= J^t_e \oplus J_s, \label{eq:T_spiking_add}\\
    J^t_e &= \Phi^t(F^t_e), \\
    F^t_e &= C_3(\Phi^t(C_{3}(\Phi^t(\Omega_{(3,5,7)}(E^t))))), \label{eq:T_fe}\\
    \Gamma(\cdot) & := \Phi^t(\Psi(\cdot)),
\end{align} 
where $\mathcal{M}$ is an operator for obtaining membrane potentials from an SNN layer, and $\Psi$ represents a fully connected layer; $\Phi^t(\cdot)$ indicates an SNN layer, which records the previous spiking status, $P^{t-1}$, and potential value, $V^{t-1}$. When receiving membrane potentials $X^t$, this SNN layer outputs updated spikes, $P^t$, and updates the recorded membrane potential $V^t$ as follows:
\begin{align}
    P^t & =h(V^t-\Theta), \nonumber\\ 
    V^t & =\alpha V^{t-1}(1-P^{t-1}) + X^t, \nonumber\\
    h(x) &= \begin{cases}1 & x>=0 \label{eq:SNN_layer} \\
0 & x<0 \end{cases},
\end{align}
where $\Theta$ is the membrane potential threshold set to 0.3 in all our experiments. The parameter $\alpha$ is a decay factor used for achieving hyperpolarization. The potential value $V^t$ is updated such that, for a spike at timestamp $t-1$, the membrane potential should be reset to $0$ by scaling $1-P^{t-1}$, and $X^t$ is the corresponding item here. 

Finally, the emotion is the average of $O^t$, $t \in [1, n]$:
\begin{align}
    R &= \sigma(\frac{1}{n}\sum_{t=1}^{n} O^{t}),
\label{eq:final_result}
\end{align}
where $\sigma$ is a Softmax activation function.
\subsection{Weight-Copy Scheme} 
Intuitively, we want temporal information extraction to focus on informative spatial positions, such as facial action units \cite{ekman1978facial}. However, events lack sufficient texture information, which impedes the temporal feature extractor from considering spatial information. To alleviate this problem, we propose a weight-copy scheme that copies the weights from the spatial feature extractor to the temporal feature extractor. Thus, during training, only the fully connected layers in $T$ are trained. The weight-copy scheme requires that all convolutional blocks before the spiking-addition operator, \ie \autoref{eq:T_spiking_add}, are of the same architecture in $S$ and $T$; see \autoref{eq:S_fs} and \autoref{eq:T_fe}. Note that the supervised loss conveys the impact from both the spatial and temporal domains enabled by the spiking-addition. Since the weights are fixed before the spiking-addition in the temporal feature extractor $T$, the training of the spatial features must also account for temporal cues. Therefore, the weight updating implicitly bridges the domain gap between intensity and event frames.

Weight copying is also applied to the self-attention weights, \ie the self-attention weights in \autoref{eq:T_fe} are replaced by the weights from \autoref{eq:omega}; see \autoref{fig:SEEN_pipeline}(a).
As we will show in our experimental results, this design is more effective than inferring the self-attention weights based on input events (row E in \autoref{tab:ablation} except E4-S0) and it yields a more efficient inference.
\subsection{Loss Function}
Because emotion recognition is a classification task, we use a regular cross-entropy loss for supervised training of SEEN:
\begin{equation}
\ell =-\frac{1}{7}\sum_{i=1}^7 y_i \log (\hat{y_i}),
\end{equation}
where  $y_i$ and $\hat{y}_i$ are the predicted $i$-th emotion's probability and corresponding ground truth probability, respectively.
\UseRawInputEncoding
\section{Dataset}
\begin{figure}[t!]
  \centering
  \begin{tabular}{c}
  \includegraphics[width=1.0\linewidth]{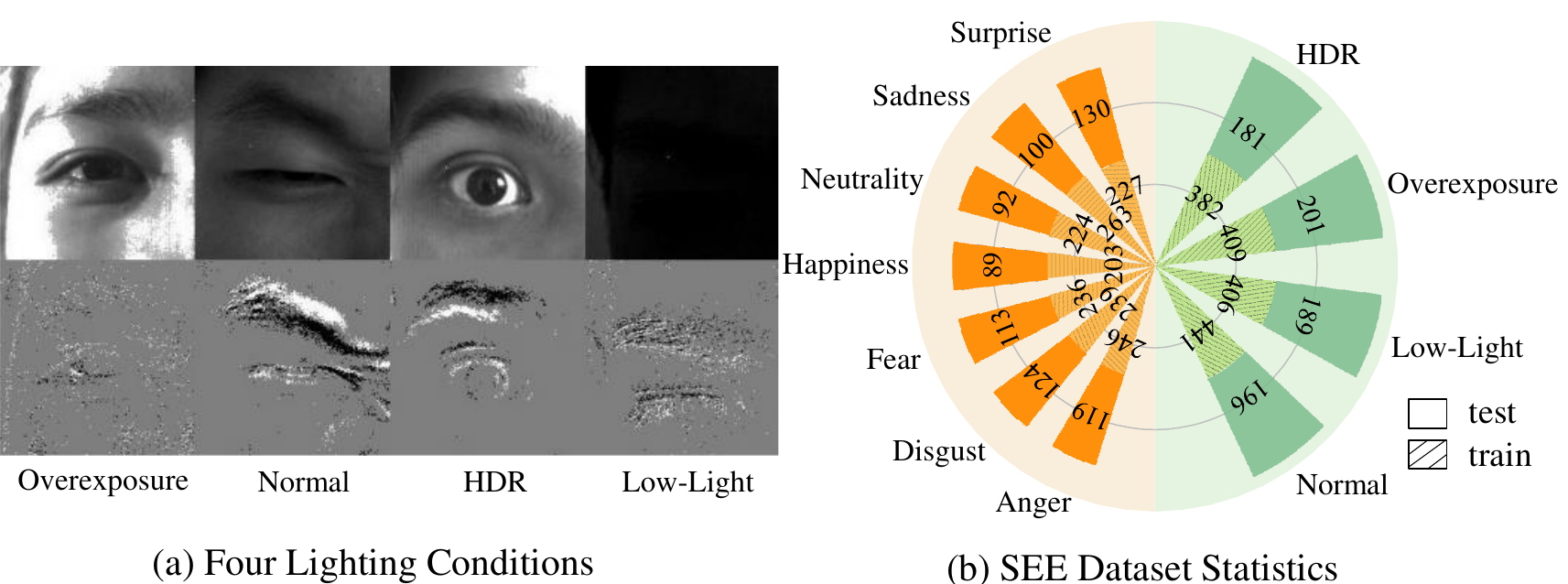}  \\	
\includegraphics[width=0.95\linewidth]{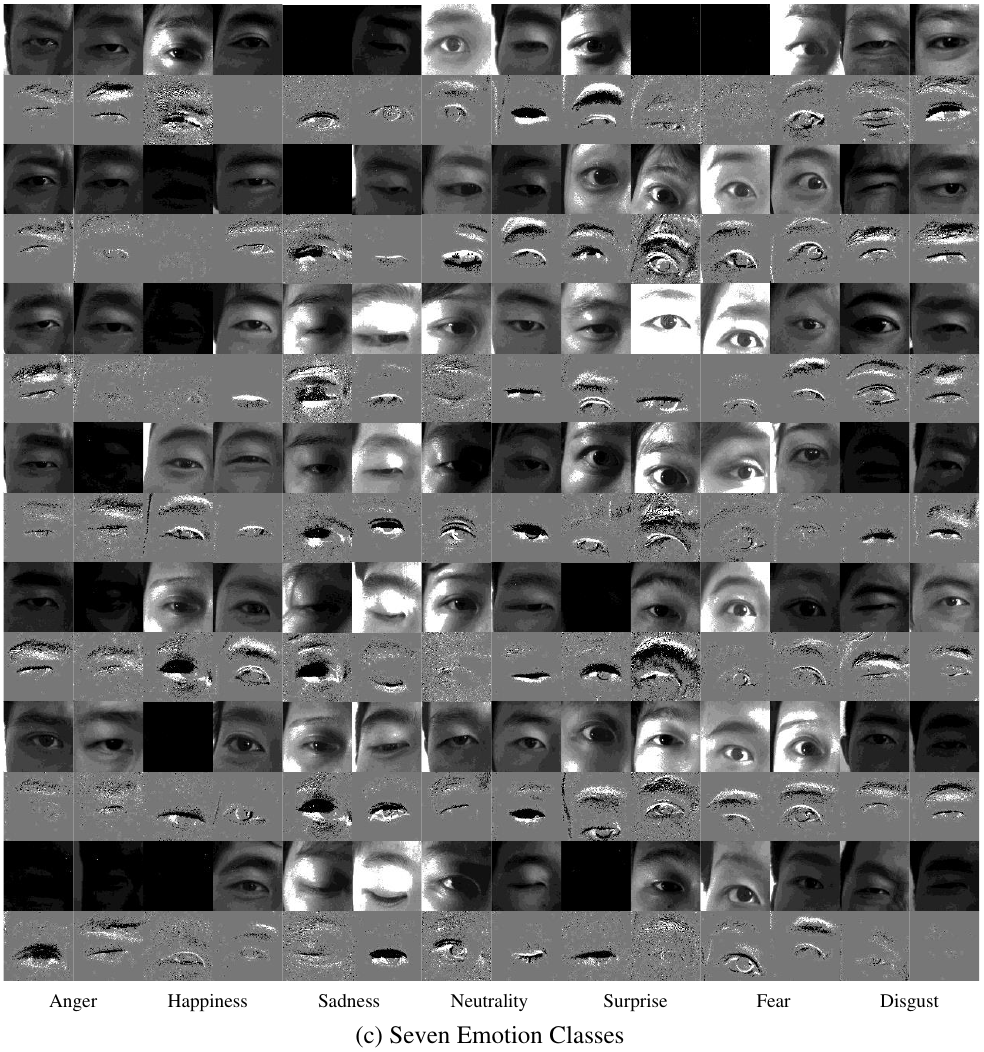} 
 \end{tabular}
  \caption{
 The newly collected Single-eye Event-based Emotion (SEE) dataset covers seven emotion classes (c) under four lighting conditions (a). The detailed statistics of the SEE dataset are illustrated in (b).
}
  \label{fig:sta_dataset}
\end{figure}

To the best of our knowledge, there does not exist an event-based dataset for single-eye emotion recognition.  The two most related are the active infrared lighting/camera datasets Eyemotion \cite{hickson2019eyemotion} (both eyes) and EMO \cite{wu2020emo} (single eye).

To address this lack of training data for event-based emotion recognition, we collect a new Single-eye Event-based Emotion (SEE) dataset; see \autoref{fig:sta_dataset}.  SEE contains data from $111$ volunteers captured with a DAVIS346 event-based camera placed in front of the right eye and mounted on a helmet; see \autoref{fig:teaser}. The DAVIS346 camera is equipped with a dynamic version sensor (DVS) and an active pixel sensor (APS), providing both raw events and conventional frames simultaneously. Unlike Eyemotion and EMO, our approach does not require any active lighting source, thereby simplifying installation, testing, and maintenance of the hardware setup. A summary of the technical differences between SEE and the existing emotion datasets is provided in Supplementary Materials. 

SEE contains videos of $7$ emotions (see~\autoref{fig:sta_dataset}(c) for an example) under four different lighting conditions: normal, overexposure, low-light, and high dynamic range (HDR) (\autoref{fig:sta_dataset}(a)).  The average video length ranges from $18$ to $131$ frames, with a mean frame number of $53.5$ and a standard deviation of $15.2$ frames, reflecting the differences in the duration of emotions between subjects. In total, SEE contains $2,405$/$128,712$ sequences/frames with corresponding raw events for a total length of $71.5$ minutes (\autoref{fig:sta_dataset}(b)), which we split in $1,638$ and $767$ sequences for training and testing, respectively.  

\UseRawInputEncoding
\section{Assessment}

The main goal of SEEN is to recognize an emotion for any phase of the emotion. Consequently, when evaluating a test sequence, we choose a uniformly distributed random starting point and corresponding testing length. 
A start point is selected such that the rest sequence is longer than the corresponding testing length.
The testing length is defined as the total accumulation time of all included event frames, $x$, and a skip time, $y$, between two adjacent event frames, denoted as E$x$-S$y$. The skip time defines a window in the time domain where all events are ignored; see ``skip'' in \autoref{fig:SEEN_pipeline}. Note that the skip time is not associated with event-based cameras but an experimental setting. Without loss of generality, the accumulation time and skip time are expressed as a multiple of $1/30$ s. Thus, E$x$-S$y$ indicates a testing length equal to $(x+ (x-1) \times y) / 30$ s. 
To reduce the impact of the randomness, we evaluate all competing methods $20$ times for different randomly selected start points for each testing sequence; we use the same random starting points for single-frame competing methods, where only the random start frame is used.
To evaluate the proposed approach and compare it to competing methods, we adopt two widely used metrics: Unweighted Average Recall (UAR) and Weighted Average Recall (WAR)~\cite{2011Cross}. UAR reflects the average accuracy of different emotion classes without considering instances per class, while WAR indicates the accuracy of overall emotions; we refer to the Supplementary Materials for formal definitions of both metrics.

\setlength{\tabcolsep}{3pt}
\begin{table*}[htbp]
\caption{ 
Quantitative comparison against the state-of-the-art. All methods are retrained and tested on the SEE dataset. The abbreviations are defined as Ha $\rightarrow$ Happiness; Sa $\rightarrow$ Sadness; An $\rightarrow$ Anger; Di $\rightarrow$ Disgust; Su $\rightarrow$ Surprise; Fe $\rightarrow$ Fear; Ne $\rightarrow$ Neutrality; Nor $\rightarrow$ Normal; Over $\rightarrow$ Overexposure; Low $\rightarrow$ Low-Light. The first and second best results are highlighted in \tb{bold} and \underline{underline}, respectively.}
	\small
	\centering
    \scalebox{1.0}{
	\begin{tabular}{l|c|ccccccc|cccc|cc|cc}
		\hline
		\hline 
  \multirow{2}{*}{Methods} & \multirow{2}{*}{} & \multicolumn{7}{c|}{Acc. of Emotion Class (\%)} & \multicolumn{4}{c|}{Acc. under Light Conditions (\%)} & \multicolumn{2}{c|}{Metrics (\%)} & \\
  \cline{3-17}
	 &  &  Ha & Sa & An  &  Di &  Su & Fe & Ne & \makebox[0.045\textwidth][c]{Nor} & \makebox[0.045\textwidth][c]{Over} & \makebox[0.045\textwidth][c]{Low} & \makebox[0.045\textwidth][c]{HDR} & WAR $\uparrow$ & UAR $\uparrow$ & FLOPS (G) & Time (ms) \\
	\hline
     Resnet18 + LSTM ~\shortcite{he2016deep, hochreiter1997long} & Face & 57.8 & \underline{86.0} & 64.9 & 46.5 & 9.2 & \underline{81.6} & 59.8 & 57.9 & 60.4 & 53.9 & 52.5 & 56.3 & 58.0 & 7.9 & 5.0\\
    Resnet50 + GRU~\shortcite{he2016deep, deng2020multitask} & Face & 27.9 & 38.0 & 49.7 & 44.5 & 6.9 & 70.0 & 5.6 & 43.0 & 35.7 & 28.9 & 32.8 & 35.2 & 34.7 & 17.3 & 10.3 \\
	 3D Resnet18~\shortcite{hara2018can} & Face & 54.8 & 45.4 & 67.7 & 23.8 & 37.2 & 42.8 & 81.6 & 51.9 & 51.4 & 44.8 & 47.8 & 49.1 & 50.5 & 8.3 & 21.2 \\
R(2+1)D~\shortcite{tran2018closer} & Face & 63.6 & 45.5 & 65.7 & 27.8 & 33.3 & 37.9 & 86.6 & 54.3 & 50.3 & 44.4 & 49.3 & 49.7 & 51.5 & 42.4 & 47.3 \\
Former DFER~\shortcite{zhao2021former} & Face & \underline{81.5} & 75.2 & \underline{85.8} & 59.4 & 39.3 & 50.8 & 78.6 & 70.1 & 65.4 & 66.2 & 61.1 & 65.8 & 67.2 & 8.3 & 7.7 \\
Former DFER w/o pre-train & Face & 44.1 & 65.2 &  46.0 & 66.5 & 28.0  & 50.3  & 36.1 & 47.0 & 51.9  & 45.6 & 47.2 & 48.0 & 48.0 & 8.3 & 7.7 \\
    \hline
Eyemotion~\shortcite{hickson2019eyemotion} & Eye & 74.3 & 85.5 & 79.5 & \underline{74.3} & \underline{69.1} & 79.2 & \underline{94.5} & \underline{79.0} & \underline{81.8} & \textbf{81.5} & \underline{72.5} & \underline{78.8} & \underline{ 79.5} & 5.7 & 17.5 \\
Eyemotion w/o pre-train & Eye & 79.6 & 85.7 & 81.2 & 71.2 & 54.7 & 71.6 & \textbf{96.4} & 77.8 & 75.9 & 79.8 & 69.7 & 75.9 & 77.2 & 5.7 & 17.5 \\
EMO~\shortcite{wu2020emo} & Eye & 75.0 & 75.1 & 70.2 & 48.1 & 37.5 & 54.1 & 82.8 & 61.8 & 62.8 & 60.1 & 69.6 & 63.1 & 63.3 & 0.3 & 7.1 \\
EMO w/o pre-train & Eye & 62.0& 73.2 & 60.1 & 38.7 & 25.7 & 48.0 & 65.3 & 46.1 & 60.2 & 55.5 & 58.9 & 53.2 & 53.3 & 0.3 & 7.1\\
    \hline   
    Ours(E$4$-S$0$) & Eye & 76.0 & 85.0 & 85.8 & 74.8 & 66.8 & 79.9 & 85.3 & 78.0 & 80.0 & 78.1 & 78.3 & 78.6 & 79.1 & 0.9 & 7.2\\
    \hline
    Ours(E$4$-S$1$) & Eye & 76.9 & 89.2 & 88.9 & 76.3 & 69.0 & 82.3 & 86.6 & 78.5 & 83.4 & 80.5 & 81.0 & 80.9 & 81.3 & 0.9 & 7.2\\
    Ours(E$7$-S$0$) & Eye & 76.7 & 86.8 & 87.6 & 74.2 & 66.2 & 82.4 & 86.7 & 78.1 & 80.9 & 77.3 & 82.1 & 79.6 & 80.1 & 1.5 & 10.7 \\
\hline
    Ours(E$4$-S$3$) & Eye & \textbf{ 85.0} & 89.9 & \textbf{92.2} & 76.7 & \textbf{ 72.1} & \textbf{87.7} & 85.2 & \textbf{83.3} & 85.6 & \underline{80.8} & \textbf{84.8} & \textbf{83.6} & \textbf{84.1} & 0.9 & 7.2 \\
    Ours(E$7$-S$1$) & Eye & 79.0 & \textbf{90.9} & 91.1 & 77.2 & 71.7 & 85.0 & 84.4 & 82.4 & \textbf{86.7} & 79.8 & 80.3 & 82.4 & 82.7 & 1.5 & 10.7 \\
    Ours(E${13}$-S$0$) & Eye  & 77.9 & 88.7 & 90.2 & \textbf{79.2} & 69.7 & 87.6 & 84.6 & 81.1 & 86.5 & 79.4 & 81.8 & 82.3 & 82.5 & 2.6 & 19.0 \\
    \hline
	\hline
	\end{tabular}}
	\label{tab:overall}
\end{table*}

\subsection{Training Setup} 
SEEN is implemented in PyTorch \cite{NEURIPS2019_9015} and trained with stochastic gradient descent (SGD) with a momentum of $0.9$ and a weight decay of $1\mathrm{e}{-3}$. We train SEEN for $180$ epochs with a batch size of $32$ on an NVIDIA TITAN V GPU. We use the StepLR scheduler to moderate the learning rate. Specifically, the initial learning rate is set to $0.015$, the step size is set to $1$, and the decay rate is set to $0.94$. 
For the SNN settings, we use a spiking threshold of $0.3$ and a decay factor of $0.2$ for all SNN neurons.

\subsection{Qualitative and Quantitative Evaluation} 
We compare the effectiveness of SEEN to existing emotion recognition methods relying on conventional intensity images only, including whole-face, single-eye, and double-eye based methods. Of these prior methods, Eyemotion \cite{hickson2019eyemotion} and EMO \cite{wu2020emo} are single-frame methods for predicting an emotion, while all other methods require the full video sequence. As shown in \autoref{tab:overall}, SEEN for E$4$-S$3$ offers the best performance, outperforming the runner-up method, Eyemotion, by significant margins, $4.8$\% and $4.6$\% higher in WAR and UAR, respectively. Under normal, overexposure, and HDR lighting conditions, our approach with the same setting also outperforms Eyemotion by at least $4\%$ in accuracy. 
However, Eyemotion offers slightly better performance under low-light conditions than SEEN with E$4$-S$3$. We posit that Eymotion benefits from the Imagenet\cite{5206848} pre-training process; without this pre-training step, Eyemotion's accuracy is $1\%$ less than the one offered by SEEN with E$4$-S$3$ setting. Moreover, we note that Eyemotion requires a personalization preprocessing step, which requires subtracting a mean neutral image for each person. Personalization dramatically increases the accuracy of neutral emotion estimation regardless of whether Eyemotion is pre-trained on ImageNet or not.

We compare SEEN with three different sequence lengths: $4/30$ s, \ie E$4$-S$0$; $7/30$ s, \ie E$4$-S$1$ group; $13/30$ s, \ie E$4$-S$3$ group. The experimental results show that the accuracy of SEEN improves with longer sequence length under all lighting conditions, especially under HDR conditions. Note, all other prior video-based approaches require the full video sequences; consequently, their delay time is the length of an input sequence. In contrast, our method can flexibly adjust the delay time by changing input settings. 
Figures \ref{fig:visual_1} and \ref{fig:visual_2} qualitatively demonstrate the benefits of our method compared to prior eye-based emotion recognition methods.
In \autoref{tab:overall}, the complexity and processing speed of each competing approach are also provided. As the temporal feature extractor processes event frames iteratively, the complexity and processing time increase with the number of event frames. Nevertheless, with the E$4$-S$3$ setting, our method offers the second fastest processing speed, but it is more than $20$\% more accurate than the fastest method, EMO.

\setlength{\tabcolsep}{2.0pt}
\begin{table}[htbp]
    \caption{
    Quantitative ablation comparisons show that: a) both the first and last intensity frames are essential for providing discriminative features; b) all components of SEEN contribute to the overall performance (except experiment E under the E$4$-S$0$ setting); and c) potential averaging is necessary results in a more accurate performance. 
    }
	\centering
    \scalebox{0.95}{
	\begin{tabular}{cl|cc|cc|cc}
		\hline
		\hline
   & & \multicolumn{2}{c|}{E$4$-S$0$} & \multicolumn{2}{c|}{E$4$-S$1$} & \multicolumn{2}{c}{E${4}$-S$3$} \\
   \hline
    & Networks &WAR   &UAR  &WAR &UAR    &WAR &UAR   \\

    \hline
    $A$	& w/o $I^n$ & 77.1 & 77.6  & 79.9 & 80.2 & 81.3 & 81.8 \\  
    $B$	& $I^n \rightarrow$ $I^2$  & 76.4 & 76.9 & 80.1 & 80.6  & 81.8 & 82.2 \\  
    $C$ & $[I^1,...,I^n]$  & 78.0 & 78.4 & 79.9 & 80.2 & 82.9& 83.3\\ 
    \hline
    \hline
    $D$	& No weight copy & 77.5 & 78.0 & 79.6 & 80.0 & 82.1 & 82.6 \\ 
    $E$	& No Att. weight copy & 78.7 & 79.2 & 80.7 & 81.1 & 83.0 & 83.2 \\ 

    \hline
    $F$	& SNN $\rightarrow$ CNN  & 50.2 & 50.2 & 53.2 & 53.2  & 55.7 & 55.6 \\ 
    $G$	& SNN $\rightarrow$ LSTM  & 52.9 & 53.0  & 55.3 & 55.2 & 55.8 & 55.7 \\ 
    
    $H$ & SNN $\rightarrow$ Transformer  & 69.2 & 69.8 & 73.6 & 74.2  & 77.1 & 77.3 \\ 
    $I$ & SNN $\rightarrow$ 3D CNN & 54.3 & 54.3  & 57.7 & 57.7  & 59.9 & 59.9 \\ 
    \hline
    \hline
    $J$	& Last potential  & 76.6 & 77.2 & 78.8 & 79.2 & 81.1 & 81.7  \\
    $K$	& Last spike  & 55.7 & 54.8  & 59.5 & 58.9  & 63.2 & 62.8  \\ 
    $L$	& Mean spike  & 63.5 & 63.2 & 64.1 & 63.6  & 69.7 & 69.5 \\ 
    \hline
    \hline

    $M$	& Ours & 78.6 & 79.1 & 80.9 & 81.3 & 83.6 & 84.1 \\ 

	\hline
	\hline
	\end{tabular}
	}	\label{tab:ablation}
\end{table}

\subsection{Ablation Study}
To gain better insight into the abilities of SEEN, we perform a series of ablation studies that investigate a) the impacts of input, b) the influence of each component of SEEN, and c) the impact of outputs. \autoref{tab:ablation} summarizes the experimental results.

\textit{Impacts of Input.} SEEN leverages the first and last intensity frames. Experiments (A), (B) and (C) gauge the impact of the intensity frames: experiment (A) only uses the first intensity frame, experiment (B) replaces the last intensity frame with the second frame, and experiment (C) uses all the intensity frames corresponding to the included event frames. The results of (A) and (B) demonstrate spatial differences are critical for $T$ to extract descriptive temporal cues.  
Compared to experiments (A) and (B), the results of experiment (C) show that using more intensity frames slightly increases performance. However, compared to our method, the setup dramatically increases data bandwidth. 

\textit{Influence of SEEN components.} We investigate the effectiveness of the different components that comprise SEEN: 1) the effectiveness of 
the weight-copy scheme (experiments (D) and (E)) and 2) the benefits of SNNs (experiments (F) to (I)). These two experiment groups show that SEEN with all components offers the best performance, except experiment E under the E$4$-S$0$ setting.
Experiments (F) to (I) show that replacing the CNN-SNN with a 3-layer CNN, LSTM, Transformer, or 3D CNN significantly degrades performance. A CNN fails to extract useful temporal cues, so the performance degradation further justifies the inclusion of temporal cues. Although LSTM, Transformer, and 3D CNN can extract temporal cues, they are less effective than SNNs. Notably, an SNN neuron's spiking mechanism acts as temporal memory and a natural noise filter, which is beneficial for robust emotion recognition. 

\textit{Impact of outputs.}
SEEN estimates emotions based on the average of $n$ membrane potentials; see \autoref{eq:T_O} and \autoref{eq:final_result}. To better understand the impact of this design decision, we conduct three ablation experiments: instead of using the average of $n$ membrane potentials, we define the prediction score based on the potential generated by the last event frame only (experiment (J)); similar to the previous but using output spikes instead of potential (experiment (K)); and finally using the average of $n$ output spikes instead of the $n$ membrane potentials for emotion classification, \ie remove the $\mathcal{M}$ operator in \autoref{eq:T_O} (experiment (L)). These results show that membrane potentials are more effective signals than spikes. We posit that the higher precision of membrane potentials (float vs. binary for spikes) offers more discriminative features for emotion classification. 
When a membrane potential triggers a spike, the potential is reset to $0$. However, it becomes a problem if we leverage the potential as an output signal since the rest operation breaks the temporal cues. To address the problem, we design to use the average of the output potentials as the output signal. Experiment (J) validates the effectiveness of this design.

\UseRawInputEncoding
\section{Conclusion}
In this work, we introduce a novel wearable single-eye-based emotion recognition prototype that can effectively estimate emotions under challenging lighting conditions. To this end, we investigate event-based camera inputs for emotion recognition. Due to the high dynamic range and temporal resolution of event-based cameras, the captured events can robustly encode temporal information under different lighting conditions. However, the captured events are asynchronous, noisy, and lack texture cues. We introduce SEEN, a novel learning-based solution to  extract informative temporal cues for emotion recognition. SEEN introduces two novel design components: a weight-copy scheme and a CNN-SNN-based temporal feature extractor. The former injects spatial attention into temporal feature extraction during the training and inference phases. The latter exploits both spatial awareness and the spiking mechanism of SNNs to provide discriminative features for emotion classification effectively. Our extensive experimental results show that SEEN can effectively estimate an emotion  from any phase of the emotion. To the best of our knowledge, SEEN is the first attempt at leveraging event-based cameras and SNNs for  emotion recognition tasks. 

\begin{acks}
This work was supported in part by the National Key Research and Development Program of China (2022ZD0\-210500), the National Natural Science Foundation of China under Grants  61972067/U21A2049-1, and the Distinguished Young Scholars Funding of Dalian (No. 2022RJ01).  Pieter Peers was supported in part by NSF grant IIS-1909028. Felix Heide was supported by an NSF CAREER Award (2047359), a Packard Foundation Fellowship, a Sloan Research Fellowship, a Sony Young Faculty Award, a Project X Innovation Award, and an Amazon Science Research Award.
\end{acks}
 
\bibliographystyle{ACM-Reference-Format}
\bibliography{sample-base}


\begin{thebibliography}{45}


\ifx \showCODEN    \undefined \def \showCODEN     #1{\unskip}     \fi
\ifx \showDOI      \undefined \def \showDOI       #1{#1}\fi
\ifx \showISBNx    \undefined \def \showISBNx     #1{\unskip}     \fi
\ifx \showISBNxiii \undefined \def \showISBNxiii  #1{\unskip}     \fi
\ifx \showISSN     \undefined \def \showISSN      #1{\unskip}     \fi
\ifx \showLCCN     \undefined \def \showLCCN      #1{\unskip}     \fi
\ifx \shownote     \undefined \def \shownote      #1{#1}          \fi
\ifx \showarticletitle \undefined \def \showarticletitle #1{#1}   \fi
\ifx \showURL      \undefined \def \showURL       {\relax}        \fi
\providecommand\bibfield[2]{#2}
\providecommand\bibinfo[2]{#2}
\providecommand\natexlab[1]{#1}
\providecommand\showeprint[2][]{arXiv:#2}

\bibitem[Appelhans and Luecken(2006)]%
        {doi:10.1037/1089-2680.10.3.229}
\bibfield{author}{\bibinfo{person}{Bradley~M. Appelhans} {and} \bibinfo{person}{Linda~J. Luecken}.} \bibinfo{year}{2006}\natexlab{}.
\newblock \showarticletitle{Heart Rate Variability as an Index of Regulated Emotional Responding}.
\newblock \bibinfo{journal}{\emph{Review of General Psychology}} \bibinfo{volume}{10}, \bibinfo{number}{3} (\bibinfo{year}{2006}), \bibinfo{pages}{229--240}.
\newblock
\urldef\tempurl%
\url{https://doi.org/10.1037/1089-2680.10.3.229}
\showDOI{\tempurl}


\bibitem[Burgos-Artizzu et~al\mbox{.}(2015)]%
        {10.1145/2820903.2820910}
\bibfield{author}{\bibinfo{person}{Xavier~P. Burgos-Artizzu}, \bibinfo{person}{Julien Fleureau}, \bibinfo{person}{Olivier Dumas}, \bibinfo{person}{Thierry Tapie}, \bibinfo{person}{Fran\c{c}ois LeClerc}, {and} \bibinfo{person}{Nicolas Mollet}.} \bibinfo{year}{2015}\natexlab{}.
\newblock \showarticletitle{Real-Time Expression-Sensitive HMD Face Reconstruction}. In \bibinfo{booktitle}{\emph{SIGGRAPH Asia 2015 Technical Briefs}} (Kobe, Japan) \emph{(\bibinfo{series}{SA '15})}. \bibinfo{publisher}{Association for Computing Machinery}, \bibinfo{address}{New York, NY, USA}, Article \bibinfo{articleno}{9}, \bibinfo{numpages}{4}~pages.
\newblock
\showISBNx{9781450339308}
\urldef\tempurl%
\url{https://doi.org/10.1145/2820903.2820910}
\showDOI{\tempurl}


\bibitem[Costa et~al\mbox{.}(2019)]%
        {10.1145/3328911}
\bibfield{author}{\bibinfo{person}{Jean Costa}, \bibinfo{person}{Fran\c{c}ois Guimbreti\`{e}re}, \bibinfo{person}{Malte~F. Jung}, {and} \bibinfo{person}{Tanzeem Choudhury}.} \bibinfo{year}{2019}\natexlab{}.
\newblock \showarticletitle{BoostMeUp: Improving Cognitive Performance in the Moment by Unobtrusively Regulating Emotions with a Smartwatch}.
\newblock \bibinfo{journal}{\emph{Proc. ACM Interact. Mob. Wearable Ubiquitous Technol.}} \bibinfo{volume}{3}, \bibinfo{number}{2}, Article \bibinfo{articleno}{40} (\bibinfo{date}{jun} \bibinfo{year}{2019}), \bibinfo{numpages}{23}~pages.
\newblock
\urldef\tempurl%
\url{https://doi.org/10.1145/3328911}
\showDOI{\tempurl}


\bibitem[Couret et~al\mbox{.}(2019)]%
        {Couret2019TheEO}
\bibfield{author}{\bibinfo{person}{David Couret}, \bibinfo{person}{Pierre Simeone}, \bibinfo{person}{S{\'e}bastien Freppel}, {and} \bibinfo{person}{Lionel~J Velly}.} \bibinfo{year}{2019}\natexlab{}.
\newblock \showarticletitle{The effect of ambient-light conditions on quantitative pupillometry: a history of rubber cup}.
\newblock \bibinfo{journal}{\emph{Neurocritical Care}}  \bibinfo{volume}{30} (\bibinfo{year}{2019}), \bibinfo{pages}{492--493}.
\newblock


\bibitem[Deng et~al\mbox{.}(2020a)]%
        {deng2020multitask}
\bibfield{author}{\bibinfo{person}{Didan Deng}, \bibinfo{person}{Zhaokang Chen}, {and} \bibinfo{person}{Bertram~E Shi}.} \bibinfo{year}{2020}\natexlab{a}.
\newblock \showarticletitle{Multitask emotion recognition with incomplete labels}. In \bibinfo{booktitle}{\emph{2020 15th IEEE International Conference on Automatic Face and Gesture Recognition (FG 2020)}} (Buenos Aires, Argentina). IEEE, \bibinfo{pages}{592--599}.
\newblock
\urldef\tempurl%
\url{https://doi.org/10.1109/FG47880.2020.00131}
\showDOI{\tempurl}


\bibitem[Deng et~al\mbox{.}(2020b)]%
        {deng2020mimamo}
\bibfield{author}{\bibinfo{person}{Didan Deng}, \bibinfo{person}{Zhaokang Chen}, \bibinfo{person}{Yuqian Zhou}, {and} \bibinfo{person}{Bertram Shi}.} \bibinfo{year}{2020}\natexlab{b}.
\newblock \showarticletitle{Mimamo net: Integrating micro-and macro-motion for video emotion recognition}. In \bibinfo{booktitle}{\emph{Proceedings of the AAAI Conference on Artificial Intelligence}}, Vol.~\bibinfo{volume}{34}. Assoc Advancement Artificial Intelligence, \bibinfo{pages}{2621--2628}.
\newblock


\bibitem[Deng et~al\mbox{.}(2009)]%
        {5206848}
\bibfield{author}{\bibinfo{person}{Jia Deng}, \bibinfo{person}{Wei Dong}, \bibinfo{person}{Richard Socher}, \bibinfo{person}{Li-Jia Li}, \bibinfo{person}{Kai Li}, {and} \bibinfo{person}{Li Fei-Fei}.} \bibinfo{year}{2009}\natexlab{}.
\newblock \showarticletitle{ImageNet: A large-scale hierarchical image database}. In \bibinfo{booktitle}{\emph{2009 IEEE Conference on Computer Vision and Pattern Recognition}}. \bibinfo{pages}{248--255}.
\newblock
\urldef\tempurl%
\url{https://doi.org/10.1109/CVPR.2009.5206848}
\showDOI{\tempurl}


\bibitem[Ding et~al\mbox{.}(2022)]%
        {ding2022biologically}
\bibfield{author}{\bibinfo{person}{Jianchuan Ding}, \bibinfo{person}{Bo Dong}, \bibinfo{person}{Felix Heide}, \bibinfo{person}{Yufei Ding}, \bibinfo{person}{Yunduo Zhou}, \bibinfo{person}{Baocai Yin}, {and} \bibinfo{person}{Xin Yang}.} \bibinfo{year}{2022}\natexlab{}.
\newblock \showarticletitle{Biologically Inspired Dynamic Thresholds for Spiking Neural Networks}. In \bibinfo{booktitle}{\emph{Advances in Neural Information Processing Systems}}.
\newblock
\urldef\tempurl%
\url{https://doi.org/10.48550/arXiv.2206.04426}
\showDOI{\tempurl}


\bibitem[Ekman and Friesen(1978)]%
        {ekman1978facial}
\bibfield{author}{\bibinfo{person}{Paul Ekman} {and} \bibinfo{person}{Wallace~V Friesen}.} \bibinfo{year}{1978}\natexlab{}.
\newblock \bibinfo{booktitle}{\emph{Facial action coding systems}}.
\newblock \bibinfo{publisher}{Consulting Psychologists Press}.
\newblock


\bibitem[Gallego et~al\mbox{.}(2022)]%
        {9138762}
\bibfield{author}{\bibinfo{person}{Guillermo Gallego}, \bibinfo{person}{Tobi Delbrück}, \bibinfo{person}{Garrick Orchard}, \bibinfo{person}{Chiara Bartolozzi}, \bibinfo{person}{Brian Taba}, \bibinfo{person}{Andrea Censi}, \bibinfo{person}{Stefan Leutenegger}, \bibinfo{person}{Andrew~J. Davison}, \bibinfo{person}{Jörg Conradt}, \bibinfo{person}{Kostas Daniilidis}, {and} \bibinfo{person}{Davide Scaramuzza}.} \bibinfo{year}{2022}\natexlab{}.
\newblock \showarticletitle{Event-Based Vision: A Survey}.
\newblock \bibinfo{journal}{\emph{IEEE Transactions on Pattern Analysis and Machine Intelligence}} \bibinfo{volume}{44}, \bibinfo{number}{1} (\bibinfo{year}{2022}), \bibinfo{pages}{154--180}.
\newblock
\urldef\tempurl%
\url{https://doi.org/10.1109/TPAMI.2020.3008413}
\showDOI{\tempurl}


\bibitem[Gehrig et~al\mbox{.}(2019)]%
        {gehrig2019end}
\bibfield{author}{\bibinfo{person}{Daniel Gehrig}, \bibinfo{person}{Antonio Loquercio}, \bibinfo{person}{Konstantinos~G Derpanis}, {and} \bibinfo{person}{Davide Scaramuzza}.} \bibinfo{year}{2019}\natexlab{}.
\newblock \showarticletitle{End-to-end learning of representations for asynchronous event-based data}. In \bibinfo{booktitle}{\emph{Proceedings of the IEEE/CVF International Conference on Computer Vision}}. \bibinfo{pages}{5633--5643}.
\newblock
\urldef\tempurl%
\url{https://doi.org/10.1109/ICCV.2019.00573}
\showDOI{\tempurl}


\bibitem[Gehrig et~al\mbox{.}(2021)]%
        {gehrig2021combining}
\bibfield{author}{\bibinfo{person}{Daniel Gehrig}, \bibinfo{person}{Michelle R{\"u}egg}, \bibinfo{person}{Mathias Gehrig}, \bibinfo{person}{Javier Hidalgo-Carri{\'o}}, {and} \bibinfo{person}{Davide Scaramuzza}.} \bibinfo{year}{2021}\natexlab{}.
\newblock \showarticletitle{Combining Events and Frames Using Recurrent Asynchronous Multimodal Networks for Monocular Depth Prediction}.
\newblock \bibinfo{journal}{\emph{IEEE Robotics and Automation Letters}} \bibinfo{volume}{6}, \bibinfo{number}{2} (\bibinfo{year}{2021}), \bibinfo{pages}{2822--2829}.
\newblock
\urldef\tempurl%
\url{https://doi.org/10.1109/LRA.2021.3060707}
\showDOI{\tempurl}


\bibitem[Georgescu and Ionescu(2019)]%
        {georgescu2019recognizing}
\bibfield{author}{\bibinfo{person}{Mariana-Iuliana Georgescu} {and} \bibinfo{person}{Radu~Tudor Ionescu}.} \bibinfo{year}{2019}\natexlab{}.
\newblock \showarticletitle{Recognizing facial expressions of occluded faces using convolutional neural networks}. In \bibinfo{booktitle}{\emph{International Conference on Neural Information Processing}}, Vol.~\bibinfo{volume}{1142}. Springer, \bibinfo{pages}{645--653}.
\newblock
\urldef\tempurl%
\url{https://doi.org/10.1007/978-3-030-36808-1\_70}
\showDOI{\tempurl}


\bibitem[{Gerstner} and {Kistler}(2002)]%
        {gerstner2002spiking}
\bibfield{author}{\bibinfo{person}{Wulfram {Gerstner}} {and} \bibinfo{person}{Werner~M. {Kistler}}.} \bibinfo{year}{2002}\natexlab{}.
\newblock \bibinfo{booktitle}{\emph{Spiking Neuron Models: Single Neurons, Populations, Plasticity}}.
\newblock


\bibitem[Gruebler and Suzuki(2014)]%
        {6778017}
\bibfield{author}{\bibinfo{person}{Anna Gruebler} {and} \bibinfo{person}{Kenji Suzuki}.} \bibinfo{year}{2014}\natexlab{}.
\newblock \showarticletitle{Design of a Wearable Device for Reading Positive Expressions from Facial EMG Signals}.
\newblock \bibinfo{journal}{\emph{IEEE Transactions on Affective Computing}} \bibinfo{volume}{5}, \bibinfo{number}{3} (\bibinfo{year}{2014}), \bibinfo{pages}{227--237}.
\newblock
\urldef\tempurl%
\url{https://doi.org/10.1109/TAFFC.2014.2313557}
\showDOI{\tempurl}


\bibitem[Hara et~al\mbox{.}(2018)]%
        {hara2018can}
\bibfield{author}{\bibinfo{person}{Kensho Hara}, \bibinfo{person}{Hirokatsu Kataoka}, {and} \bibinfo{person}{Yutaka Satoh}.} \bibinfo{year}{2018}\natexlab{}.
\newblock \showarticletitle{Can spatiotemporal 3d cnns retrace the history of 2d cnns and imagenet?}. In \bibinfo{booktitle}{\emph{Proceedings of the IEEE conference on Computer Vision and Pattern Recognition}}. \bibinfo{pages}{6546--6555}.
\newblock
\urldef\tempurl%
\url{https://doi.org/10.1109/CVPR.2018.00685}
\showDOI{\tempurl}


\bibitem[He et~al\mbox{.}(2016)]%
        {he2016deep}
\bibfield{author}{\bibinfo{person}{Kaiming He}, \bibinfo{person}{Xiangyu Zhang}, \bibinfo{person}{Shaoqing Ren}, {and} \bibinfo{person}{Jian Sun}.} \bibinfo{year}{2016}\natexlab{}.
\newblock \showarticletitle{Deep residual learning for image recognition}. In \bibinfo{booktitle}{\emph{Proceedings of the IEEE conference on computer vision and pattern recognition}}. \bibinfo{pages}{770--778}.
\newblock
\urldef\tempurl%
\url{https://doi.org/10.1109/CVPR.2016.90}
\showDOI{\tempurl}


\bibitem[Hickson et~al\mbox{.}(2019)]%
        {hickson2019eyemotion}
\bibfield{author}{\bibinfo{person}{Steven Hickson}, \bibinfo{person}{Nick Dufour}, \bibinfo{person}{Avneesh Sud}, \bibinfo{person}{Vivek Kwatra}, {and} \bibinfo{person}{Irfan Essa}.} \bibinfo{year}{2019}\natexlab{}.
\newblock \showarticletitle{Eyemotion: Classifying facial expressions in VR using eye-tracking cameras}. In \bibinfo{booktitle}{\emph{2019 IEEE Winter Conference on Applications of Computer Vision (WACV)}}. IEEE, \bibinfo{pages}{1626--1635}.
\newblock
\urldef\tempurl%
\url{https://doi.org/10.1109/WACV.2019.00178}
\showDOI{\tempurl}


\bibitem[Hochreiter and Schmidhuber(1997)]%
        {hochreiter1997long}
\bibfield{author}{\bibinfo{person}{Sepp Hochreiter} {and} \bibinfo{person}{J{\"u}rgen Schmidhuber}.} \bibinfo{year}{1997}\natexlab{}.
\newblock \showarticletitle{Long short-term memory}.
\newblock \bibinfo{journal}{\emph{Neural computation}} \bibinfo{volume}{9}, \bibinfo{number}{8} (\bibinfo{year}{1997}), \bibinfo{pages}{1735--1780}.
\newblock
\urldef\tempurl%
\url{https://doi.org/10.1162/neco.1997.9.8.1735}
\showDOI{\tempurl}


\bibitem[Houshmand and Khan(2020)]%
        {houshmand2020facial}
\bibfield{author}{\bibinfo{person}{Bita Houshmand} {and} \bibinfo{person}{Naimul~Mefraz Khan}.} \bibinfo{year}{2020}\natexlab{}.
\newblock \showarticletitle{Facial expression recognition under partial occlusion from virtual reality headsets based on transfer learning}. In \bibinfo{booktitle}{\emph{2020 IEEE Sixth International Conference on Multimedia Big Data (BigMM)}}. IEEE, \bibinfo{pages}{70--75}.
\newblock
\urldef\tempurl%
\url{https://doi.org/10.1109/BigMM50055.2020.00020}
\showDOI{\tempurl}


\bibitem[Ji et~al\mbox{.}(2022)]%
        {10.1145/3528233.3530745}
\bibfield{author}{\bibinfo{person}{Xinya Ji}, \bibinfo{person}{Hang Zhou}, \bibinfo{person}{Kaisiyuan Wang}, \bibinfo{person}{Qianyi Wu}, \bibinfo{person}{Wayne Wu}, \bibinfo{person}{Feng Xu}, {and} \bibinfo{person}{Xun Cao}.} \bibinfo{year}{2022}\natexlab{}.
\newblock \showarticletitle{EAMM: One-Shot Emotional Talking Face via Audio-Based Emotion-Aware Motion Model}. In \bibinfo{booktitle}{\emph{ACM SIGGRAPH 2022 Conference Proceedings}} \emph{(\bibinfo{series}{SIGGRAPH '22})}. \bibinfo{pages}{1--10}.
\newblock
\showISBNx{9781450393379}
\urldef\tempurl%
\url{https://doi.org/10.1145/3528233.3530745}
\showDOI{\tempurl}


\bibitem[Kag and Saligrama(2021)]%
        {kag2021time}
\bibfield{author}{\bibinfo{person}{Anil Kag} {and} \bibinfo{person}{Venkatesh Saligrama}.} \bibinfo{year}{2021}\natexlab{}.
\newblock \showarticletitle{Time adaptive recurrent neural network}. In \bibinfo{booktitle}{\emph{Proceedings of the IEEE/CVF Conference on Computer Vision and Pattern Recognition (CVPR)}}. \bibinfo{pages}{15149--15158}.
\newblock
\urldef\tempurl%
\url{https://doi.org/10.1109/CVPR46437.2021.01490}
\showDOI{\tempurl}


\bibitem[Lagorce et~al\mbox{.}(2017)]%
        {lagorce2016hots}
\bibfield{author}{\bibinfo{person}{Xavier Lagorce}, \bibinfo{person}{Garrick Orchard}, \bibinfo{person}{Francesco Galluppi}, \bibinfo{person}{Bertram~E Shi}, {and} \bibinfo{person}{Ryad~B Benosman}.} \bibinfo{year}{2017}\natexlab{}.
\newblock \showarticletitle{Hots: a hierarchy of event-based time-surfaces for pattern recognition}.
\newblock \bibinfo{journal}{\emph{IEEE transactions on pattern analysis and machine intelligence}} \bibinfo{volume}{39}, \bibinfo{number}{7} (\bibinfo{year}{2017}), \bibinfo{pages}{1346--1359}.
\newblock
\urldef\tempurl%
\url{https://doi.org/10.1109/TPAMI.2016.2574707}
\showDOI{\tempurl}


\bibitem[Lee et~al\mbox{.}(2019)]%
        {lee2019context}
\bibfield{author}{\bibinfo{person}{Jiyoung Lee}, \bibinfo{person}{Seungryong Kim}, \bibinfo{person}{Sunok Kim}, \bibinfo{person}{Jungin Park}, {and} \bibinfo{person}{Kwanghoon Sohn}.} \bibinfo{year}{2019}\natexlab{}.
\newblock \showarticletitle{Context-aware emotion recognition networks}. In \bibinfo{booktitle}{\emph{Proceedings of the IEEE/CVF international conference on computer vision}}. \bibinfo{pages}{10143--10152}.
\newblock
\urldef\tempurl%
\url{https://doi.org/10.1109/ICCV.2019.01024}
\showDOI{\tempurl}


\bibitem[Lee et~al\mbox{.}(2020)]%
        {lee2020multi}
\bibfield{author}{\bibinfo{person}{Jiyoung Lee}, \bibinfo{person}{Sunok Kim}, \bibinfo{person}{Seungryong Kim}, {and} \bibinfo{person}{Kwanghoon Sohn}.} \bibinfo{year}{2020}\natexlab{}.
\newblock \showarticletitle{Multi-modal recurrent attention networks for facial expression recognition}.
\newblock \bibinfo{journal}{\emph{IEEE Transactions on Image Processing}}  \bibinfo{volume}{29} (\bibinfo{year}{2020}), \bibinfo{pages}{6977--6991}.
\newblock
\urldef\tempurl%
\url{https://doi.org/10.1109/TIP.2020.2996086}
\showDOI{\tempurl}


\bibitem[Li et~al\mbox{.}(2015)]%
        {10.1145/2766939}
\bibfield{author}{\bibinfo{person}{Hao Li}, \bibinfo{person}{Laura Trutoiu}, \bibinfo{person}{Kyle Olszewski}, \bibinfo{person}{Lingyu Wei}, \bibinfo{person}{Tristan Trutna}, \bibinfo{person}{Pei-Lun Hsieh}, \bibinfo{person}{Aaron Nicholls}, {and} \bibinfo{person}{Chongyang Ma}.} \bibinfo{year}{2015}\natexlab{}.
\newblock \showarticletitle{Facial Performance Sensing Head-Mounted Display}.
\newblock \bibinfo{journal}{\emph{ACM Trans. Graph.}} \bibinfo{volume}{34}, \bibinfo{number}{4}, Article \bibinfo{articleno}{47} (\bibinfo{date}{jul} \bibinfo{year}{2015}), \bibinfo{numpages}{9}~pages.
\newblock
\showISSN{0730-0301}
\urldef\tempurl%
\url{https://doi.org/10.1145/2766939}
\showDOI{\tempurl}


\bibitem[Li et~al\mbox{.}(2018)]%
        {Li2018EmotionRF}
\bibfield{author}{\bibinfo{person}{Mi Li}, \bibinfo{person}{Hongpei Xu}, \bibinfo{person}{Xingwang Liu}, {and} \bibinfo{person}{Shengfu Lu}.} \bibinfo{year}{2018}\natexlab{}.
\newblock \showarticletitle{Emotion recognition from multichannel EEG signals using K-nearest neighbor classification}.
\newblock \bibinfo{journal}{\emph{Technology and Health Care}}  \bibinfo{volume}{26} (\bibinfo{date}{04} \bibinfo{year}{2018}), \bibinfo{pages}{509--519}.
\newblock
\urldef\tempurl%
\url{https://doi.org/10.3233/THC-174836}
\showDOI{\tempurl}


\bibitem[Liu et~al\mbox{.}(2020)]%
        {10.3389/fnsys.2020.00043}
\bibfield{author}{\bibinfo{person}{Junxiu Liu}, \bibinfo{person}{Guopei Wu}, \bibinfo{person}{Yuling Luo}, \bibinfo{person}{Senhui Qiu}, \bibinfo{person}{Su Yang}, \bibinfo{person}{Wei Li}, {and} \bibinfo{person}{Yifei Bi}.} \bibinfo{year}{2020}\natexlab{}.
\newblock \showarticletitle{EEG-Based Emotion Classification Using a Deep Neural Network and Sparse Autoencoder}.
\newblock \bibinfo{journal}{\emph{Frontiers in Systems Neuroscience}}  \bibinfo{volume}{14} (\bibinfo{year}{2020}).
\newblock
\showISSN{1662-5137}
\urldef\tempurl%
\url{https://doi.org/10.3389/fnsys.2020.00043}
\showDOI{\tempurl}


\bibitem[Lucero and Munhall(1999)]%
        {Lucero1999AMO}
\bibfield{author}{\bibinfo{person}{Jorge~C. Lucero} {and} \bibinfo{person}{Kevin~G. Munhall}.} \bibinfo{year}{1999}\natexlab{}.
\newblock \showarticletitle{A model of facial biomechanics for speech production.}
\newblock \bibinfo{journal}{\emph{The Journal of the Acoustical Society of America}}  \bibinfo{volume}{106 5} (\bibinfo{year}{1999}), \bibinfo{pages}{2834–2842}.
\newblock
\urldef\tempurl%
\url{https://doi.org/10.1121/1.428108}
\showDOI{\tempurl}


\bibitem[Maqueda et~al\mbox{.}(2018)]%
        {maqueda2018event}
\bibfield{author}{\bibinfo{person}{Ana~I Maqueda}, \bibinfo{person}{Antonio Loquercio}, \bibinfo{person}{Guillermo Gallego}, \bibinfo{person}{Narciso Garc{\'\i}a}, {and} \bibinfo{person}{Davide Scaramuzza}.} \bibinfo{year}{2018}\natexlab{}.
\newblock \showarticletitle{Event-based vision meets deep learning on steering prediction for self-driving cars}. In \bibinfo{booktitle}{\emph{2018 IEEE/CVF Conference on Computer Vision and Pattern Recognition}}. \bibinfo{pages}{5419--5427}.
\newblock
\urldef\tempurl%
\url{https://doi.org/10.1109/CVPR.2018.00568}
\showDOI{\tempurl}


\bibitem[Mathôt(2018)]%
        {Pupillometry:2018:02}
\bibfield{author}{\bibinfo{person}{Sebastiaan Mathôt}.} \bibinfo{year}{2018}\natexlab{}.
\newblock \showarticletitle{Pupillometry: Psychology, Physiology, and Function}.
\newblock \bibinfo{journal}{\emph{Journal of Cognition}}  \bibinfo{volume}{1} (\bibinfo{date}{02} \bibinfo{year}{2018}).
\newblock
\urldef\tempurl%
\url{https://doi.org/10.5334/joc.18}
\showDOI{\tempurl}


\bibitem[Nah et~al\mbox{.}(2019)]%
        {nah2019recurrent}
\bibfield{author}{\bibinfo{person}{Seungjun Nah}, \bibinfo{person}{Sanghyun Son}, {and} \bibinfo{person}{Kyoung~Mu Lee}.} \bibinfo{year}{2019}\natexlab{}.
\newblock \showarticletitle{Recurrent neural networks with intra-frame iterations for video deblurring}. In \bibinfo{booktitle}{\emph{Proceedings of the IEEE/CVF Conference on Computer Vision and Pattern Recognition}}. \bibinfo{pages}{8094--8103}.
\newblock
\urldef\tempurl%
\url{https://doi.org/10.1109/CVPR.2019.00829}
\showDOI{\tempurl}


\bibitem[Nie et~al\mbox{.}(2020)]%
        {9097620}
\bibfield{author}{\bibinfo{person}{Jingping Nie}, \bibinfo{person}{Yigong Hu}, \bibinfo{person}{Yuanyuting Wang}, \bibinfo{person}{Stephen Xia}, {and} \bibinfo{person}{Xiaofan Jiang}.} \bibinfo{year}{2020}\natexlab{}.
\newblock \showarticletitle{SPIDERS: Low-Cost Wireless Glasses for Continuous In-Situ Bio-Signal Acquisition and Emotion Recognition}. In \bibinfo{booktitle}{\emph{2020 IEEE/ACM Fifth International Conference on Internet-of-Things Design and Implementation (IoTDI)}}. \bibinfo{pages}{27--39}.
\newblock
\urldef\tempurl%
\url{https://doi.org/10.1109/IoTDI49375.2020.00011}
\showDOI{\tempurl}


\bibitem[Paszke et~al\mbox{.}(2019)]%
        {NEURIPS2019_9015}
\bibfield{author}{\bibinfo{person}{Adam Paszke}, \bibinfo{person}{Sam Gross}, \bibinfo{person}{Francisco Massa}, \bibinfo{person}{Adam Lerer}, \bibinfo{person}{James Bradbury}, \bibinfo{person}{Gregory Chanan}, \bibinfo{person}{Trevor Killeen}, \bibinfo{person}{Zeming Lin}, \bibinfo{person}{Natalia Gimelshein}, \bibinfo{person}{Luca Antiga}, \bibinfo{person}{Alban Desmaison}, \bibinfo{person}{Andreas Kopf}, \bibinfo{person}{Edward Yang}, \bibinfo{person}{Zachary DeVito}, \bibinfo{person}{Martin Raison}, \bibinfo{person}{Alykhan Tejani}, \bibinfo{person}{Sasank Chilamkurthy}, \bibinfo{person}{Benoit Steiner}, \bibinfo{person}{Lu Fang}, \bibinfo{person}{Junjie Bai}, {and} \bibinfo{person}{Soumith Chintala}.} \bibinfo{year}{2019}\natexlab{}.
\newblock \showarticletitle{PyTorch: An Imperative Style, High-Performance Deep Learning Library}.
\newblock In \bibinfo{booktitle}{\emph{Advances in Neural Information Processing Systems 32}}. \bibinfo{publisher}{Curran Associates, Inc.}, \bibinfo{pages}{8024--8035}.
\newblock
\urldef\tempurl%
\url{http://papers.neurips.cc/paper/9015-pytorch-an-imperative-style-high-performance-deep-learning-library.pdf}
\showURL{%
\tempurl}


\bibitem[Ruan et~al\mbox{.}(2021)]%
        {ruan2021feature}
\bibfield{author}{\bibinfo{person}{Delian Ruan}, \bibinfo{person}{Yan Yan}, \bibinfo{person}{Shenqi Lai}, \bibinfo{person}{Zhenhua Chai}, \bibinfo{person}{Chunhua Shen}, {and} \bibinfo{person}{Hanzi Wang}.} \bibinfo{year}{2021}\natexlab{}.
\newblock \showarticletitle{Feature decomposition and reconstruction learning for effective facial expression recognition}. In \bibinfo{booktitle}{\emph{Proceedings of the IEEE/CVF Conference on Computer Vision and Pattern Recognition}}. \bibinfo{pages}{7660--7669}.
\newblock
\urldef\tempurl%
\url{https://doi.org/10.1109/CVPR46437.2021.00757}
\showDOI{\tempurl}


\bibitem[Sanchez et~al\mbox{.}(2021)]%
        {sanchez2021affective}
\bibfield{author}{\bibinfo{person}{Enrique Sanchez}, \bibinfo{person}{Mani~Kumar Tellamekala}, \bibinfo{person}{Michel Valstar}, {and} \bibinfo{person}{Georgios Tzimiropoulos}.} \bibinfo{year}{2021}\natexlab{}.
\newblock \showarticletitle{Affective Processes: stochastic modelling of temporal context for emotion and facial expression recognition}. In \bibinfo{booktitle}{\emph{Proceedings of the IEEE/CVF Conference on Computer Vision and Pattern Recognition}}. \bibinfo{pages}{9074--9084}.
\newblock
\urldef\tempurl%
\url{https://doi.org/10.1109/CVPR46437.2021.00896}
\showDOI{\tempurl}


\bibitem[Schuller et~al\mbox{.}(2011)]%
        {2011Cross}
\bibfield{author}{\bibinfo{person}{B. Schuller}, \bibinfo{person}{B. Vlasenko}, \bibinfo{person}{F. Eyben}, \bibinfo{person}{M. Wo?Llmer}, \bibinfo{person}{A. Stuhlsatz}, \bibinfo{person}{A. Wendemuth}, {and} \bibinfo{person}{G. Rigoll}.} \bibinfo{year}{2011}\natexlab{}.
\newblock \showarticletitle{Cross-Corpus Acoustic Emotion Recognition: Variances and Strategies}.
\newblock \bibinfo{journal}{\emph{IEEE Transactions on Affective Computing}} \bibinfo{volume}{1}, \bibinfo{number}{2} (\bibinfo{year}{2011}), \bibinfo{pages}{119--131}.
\newblock
\urldef\tempurl%
\url{https://doi.org/10.1109/T-AFFC.2010.8}
\showDOI{\tempurl}


\bibitem[Tran et~al\mbox{.}(2018)]%
        {tran2018closer}
\bibfield{author}{\bibinfo{person}{Du Tran}, \bibinfo{person}{Heng Wang}, \bibinfo{person}{Lorenzo Torresani}, \bibinfo{person}{Jamie Ray}, \bibinfo{person}{Yann LeCun}, {and} \bibinfo{person}{Manohar Paluri}.} \bibinfo{year}{2018}\natexlab{}.
\newblock \showarticletitle{A closer look at spatiotemporal convolutions for action recognition}. In \bibinfo{booktitle}{\emph{Proceedings of the IEEE conference on Computer Vision and Pattern Recognition}}. \bibinfo{pages}{6450--6459}.
\newblock
\urldef\tempurl%
\url{https://doi.org/10.1109/CVPR.2018.00675}
\showDOI{\tempurl}


\bibitem[Wang et~al\mbox{.}(2022)]%
        {wang2021event}
\bibfield{author}{\bibinfo{person}{Yanxiang Wang}, \bibinfo{person}{Xian Zhang}, \bibinfo{person}{Yiran Shen}, \bibinfo{person}{Bowen Du}, \bibinfo{person}{Guangrong Zhao}, \bibinfo{person}{Lizhen Cui~Cui Lizhen}, {and} \bibinfo{person}{Hongkai Wen}.} \bibinfo{year}{2022}\natexlab{}.
\newblock \showarticletitle{Event-Stream Representation for Human Gaits Identification Using Deep Neural Networks}.
\newblock \bibinfo{journal}{\emph{IEEE Transactions on Pattern Analysis and Machine Intelligence}} \bibinfo{volume}{44}, \bibinfo{number}{7} (\bibinfo{year}{2022}), \bibinfo{pages}{3436--3449}.
\newblock
\urldef\tempurl%
\url{https://doi.org/10.1109/TPAMI.2021.3054886}
\showDOI{\tempurl}


\bibitem[Wu et~al\mbox{.}(2020)]%
        {wu2020emo}
\bibfield{author}{\bibinfo{person}{Hao Wu}, \bibinfo{person}{Jinghao Feng}, \bibinfo{person}{Xuejin Tian}, \bibinfo{person}{Edward Sun}, \bibinfo{person}{Yunxin Liu}, \bibinfo{person}{Bo Dong}, \bibinfo{person}{Fengyuan Xu}, {and} \bibinfo{person}{Sheng Zhong}.} \bibinfo{year}{2020}\natexlab{}.
\newblock \showarticletitle{EMO: Real-time emotion recognition from single-eye images for resource-constrained eyewear devices}. In \bibinfo{booktitle}{\emph{Proceedings of the 18th International Conference on Mobile Systems, Applications, and Services}}. \bibinfo{pages}{448--461}.
\newblock
\urldef\tempurl%
\url{https://doi.org/10.1145/3386901.3388917}
\showDOI{\tempurl}


\bibitem[Wu et~al\mbox{.}(2018)]%
        {wu2018spatio}
\bibfield{author}{\bibinfo{person}{Yujie Wu}, \bibinfo{person}{Lei Deng}, \bibinfo{person}{Guoqi Li}, \bibinfo{person}{Jun Zhu}, {and} \bibinfo{person}{Luping Shi}.} \bibinfo{year}{2018}\natexlab{}.
\newblock \showarticletitle{Spatio-temporal backpropagation for training high-performance spiking neural networks}.
\newblock \bibinfo{journal}{\emph{Frontiers in neuroscience}}  \bibinfo{volume}{12} (\bibinfo{year}{2018}), \bibinfo{pages}{331}.
\newblock
\urldef\tempurl%
\url{https://doi.org/10.3389/fnins.2018.00331}
\showDOI{\tempurl}


\bibitem[Xue et~al\mbox{.}(2021)]%
        {xue2021transfer}
\bibfield{author}{\bibinfo{person}{Fanglei Xue}, \bibinfo{person}{Qiangchang Wang}, {and} \bibinfo{person}{Guodong Guo}.} \bibinfo{year}{2021}\natexlab{}.
\newblock \showarticletitle{Transfer: Learning relation-aware facial expression representations with transformers}. In \bibinfo{booktitle}{\emph{Proceedings of the IEEE/CVF International Conference on Computer Vision}}. \bibinfo{pages}{3601--3610}.
\newblock
\urldef\tempurl%
\url{https://doi.org/10.1109/ICCV48922.2021.00358}
\showDOI{\tempurl}


\bibitem[Zhang et~al\mbox{.}(2021b)]%
        {zhang2021object}
\bibfield{author}{\bibinfo{person}{Jiqing Zhang}, \bibinfo{person}{Xin Yang}, \bibinfo{person}{Yingkai Fu}, \bibinfo{person}{Xiaopeng Wei}, \bibinfo{person}{Baocai Yin}, {and} \bibinfo{person}{Bo Dong}.} \bibinfo{year}{2021}\natexlab{b}.
\newblock \showarticletitle{Object tracking by jointly exploiting frame and event domain}. In \bibinfo{booktitle}{\emph{Proceedings of the IEEE/CVF International Conference on Computer Vision}}. \bibinfo{pages}{13043--13052}.
\newblock
\urldef\tempurl%
\url{https://doi.org/10.1109/ICCV48922.2021.01280}
\showDOI{\tempurl}


\bibitem[Zhang et~al\mbox{.}(2021a)]%
        {zhang2021relative}
\bibfield{author}{\bibinfo{person}{Yuhang Zhang}, \bibinfo{person}{Chengrui Wang}, {and} \bibinfo{person}{Weihong Deng}.} \bibinfo{year}{2021}\natexlab{a}.
\newblock \showarticletitle{Relative Uncertainty Learning for Facial Expression Recognition}.
\newblock \bibinfo{journal}{\emph{Advances in Neural Information Processing Systems}}  \bibinfo{volume}{34} (\bibinfo{year}{2021}), \bibinfo{pages}{17616--17627}.
\newblock


\bibitem[Zhao and Liu(2021)]%
        {zhao2021former}
\bibfield{author}{\bibinfo{person}{Zengqun Zhao} {and} \bibinfo{person}{Qingshan Liu}.} \bibinfo{year}{2021}\natexlab{}.
\newblock \showarticletitle{Former-DFER: Dynamic Facial Expression Recognition Transformer}. In \bibinfo{booktitle}{\emph{Proceedings of the 29th ACM International Conference on Multimedia}}. \bibinfo{pages}{1553--1561}.
\newblock
\urldef\tempurl%
\url{https://doi.org/10.1145/3474085.3475292}
\showDOI{\tempurl}


\end{thebibliography}

\UseRawInputEncoding
\appendix

\begin{figure*}[h]
  \centering
  \scalebox{1.0}{
  \begin{tabular}{c}
 \includegraphics[width=0.91\linewidth]{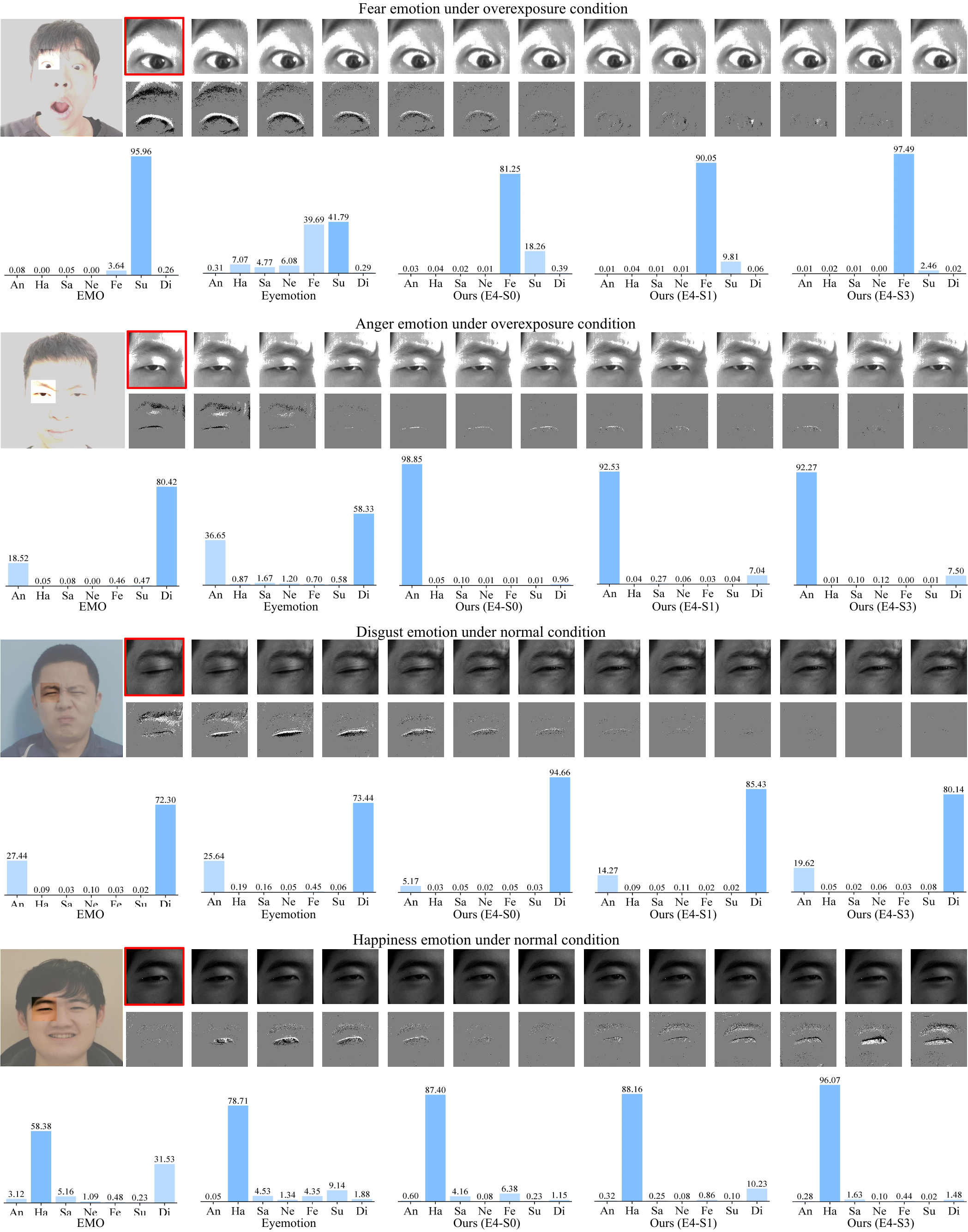}   \\

\end{tabular} }
  \caption{We show four examples across four different emotions, Fear, Anger, Disgust, and Happiness, under overexposure and normal lighting conditions. The frames marked with \rb{red} boxes are the inputs for EMO \cite{wu2020emo} and Eyemotion \cite{hickson2019eyemotion}, which is also the first input frame of our approach. Our approach offers the most accurate emotion predictions under all test settings.}
  \label{fig:visual_1}
\end{figure*}

\begin{figure*}[h]
  \centering
  \scalebox{1.0}{
  \begin{tabular}{c}
 \includegraphics[width=0.91\linewidth]{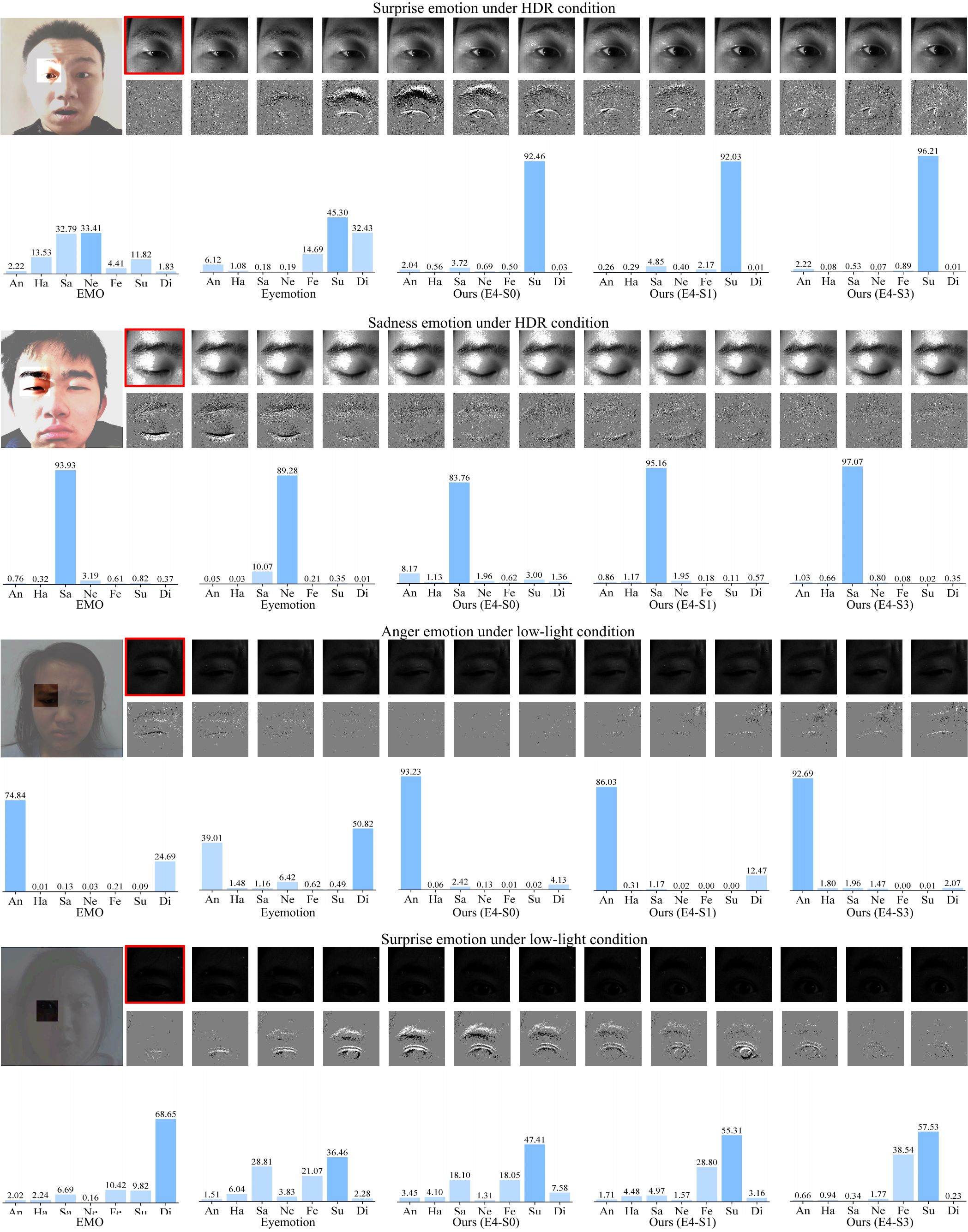}   \\

\end{tabular} }
  \caption{We show four additional examples across another four different emotions, Surprise, Sadness, Anger, and Surprise, under HDR and low-light conditions. The frames marked with \rb{red} boxes are the inputs for EMO \cite{wu2020emo} and Eyemotion \cite{hickson2019eyemotion}, which is also the first input frame of our approach. Our approach offers the most accurate emotion predictions under all test settings.}
  \label{fig:visual_2}
\end{figure*}

\end{document}